\documentclass[11pt,epsf,letterpaper]{article}
\usepackage{setspace}
\usepackage{color}
\usepackage{amsmath}
\usepackage{amsfonts}
\usepackage{verbatim}
\usepackage{amssymb}
\usepackage{graphicx}

\providecommand{\U}[1]{\protect\rule{.1in}{.1in}}
\begin{document}

\title{Scalar field perturbations in asymptotically Lifshitz black holes}
\author{Alex Giacomini$^{1}$, Gaston Giribet$^{2}$, Mauricio Leston$^{3}$,
Julio Oliva$^{1}$, Sourya Ray$^{1}$ \\
%EndAName
\textit{$^1$Instituto de Ciencias F\'{\i}sicas y Matem\'aticas, Universidad
Austral de Chile, Valdivia, Chile.}\\
\textit{$^2$Departamento de F\'{\i}sica, Universidad de Buenos Aires
FCEN-UBA y Conicet, Argentina.}\\
\textit{$^3$Instituto de Astronom\'{\i}a y F\'{\i}sica del Espacio IAFE,
UBA-Conicet, Argentina.} }
\maketitle

\begin{abstract}
We consider scalar field perturbations about asymptotically Lifshitz black
holes with dynamical exponent $z$ in $D$ dimensions. We show that, for
suitable boundary conditions, these Lifshitz black holes are stable under
scalar field perturbations. For $z=2$, we explicitly compute the quasinormal
mode frequencies, which result to be purely imaginary, and then obtain the
damping-off of the scalar field perturbation in these backgrounds. The
general analysis includes, in particular, the $z=3$ black hole solution of
three-dimensional massive gravity.
\end{abstract}

%\pacs{}

%\maketitle

\section{Introduction}

\subsection{Lifshitz holography}

It has been conjectured in \cite{Kachru} that the non-relativistic
Lifshitz-like fixed points admit a dual description in terms of holographic
gravity models, mimicking what happens in the standard AdS/CFT
correspondence for the case of relativistic conformal field theories \cite%
{Malda}. The proposal in \cite{Kachru} is part of a more general program,
whose goal is to generalize AdS/CFT correspondence to the area of condensed
matter physics. The importance of having a tool to describe
strongly-correlated condensed matter systems is undoubtable, and this is the
reason why much attention has been focused on this area in the last four
years.

In this context, a Lifshitz fixed point refers to a non-relativistic system
described by a field theory exhibiting anisotropic scale invariance under
the transformation%
\begin{equation}
\vec{x}\rightarrow\lambda^{2}\ \vec{x}\ ,\qquad t\rightarrow\lambda ^{2z}\ t
\label{Gsymmetry}
\end{equation}
with $z\in\mathbb{R}_{>1}$ in the spacetime coordinates, but no Galilean
invariance. Systems with such a behavior appear, for instance, in the
description of strongly correlated electrons. These systems are
characterized by the value of the dynamical exponent $z$ that parameterizes
the scale transformation (\ref{Gsymmetry}). The proposal in \cite{Kachru} is
that, analogously to the case of relativistic conformal field theory, the ($%
D-1$)-dimensional Lifshitz fixed points would also admit a geometric dual
description in $D$ dimensions. The geometric realization is such that the $D$%
-dimensional asymptotically Anti-de Sitter spaces of the standard AdS$_{D}$%
/CFT$_{D-1}$ picture get replaced by spaces whose metrics asymptote the
geometry%
\begin{equation}
ds^{2}=-\frac{r^{2z}}{l^{2z}}dt^{2}+\frac{l^{2}}{r^{2}}dr^{2}+r^{2}\ d\vec {x%
}_{D-2}^{2}  \label{Gmetric}
\end{equation}
at large $r$, where $d\vec{x}_{D-2}^{2}$ is the metric of the ($D-2$%
)-dimensional plane $\mathbb{R}^{D-2}$, $l$ is a positive parameter with
dimensions of length, and $z$ is a real parameter that ultimately coincides
with the dynamical exponent of the scale invariant theory that one is trying
to holographically describe. Then, spaces (\ref{Gmetric}), dubbed ``Lifshitz
spaces'', happen to realize geometrically the scaling symmetry (\ref%
{Gsymmetry}), which is certainly a symmetry of the metric above provided one
also rescales the radial coordinate as $r\rightarrow\lambda^{-2}\ r$. In
this generalized framework, the standard AdS$_{D}$/CFT$_{D-1}$
correspondence is recovered in the particular case $z=1$, for which (\ref%
{Gmetric}) coincides with AdS$_{D}$ in Poincar\'{e} coordinates. This
suggests a natural way of prescribing the holographic dictionary by
mimicking the recipe of \cite{Witten, GKP} and adapting it to generic values
of $z$. This holographic correspondence was thoughtfully studied, for
instance, in references \cite{Ross, Ross2, Horava, manguero1, manguero2,
manguero3}. As seen in \cite{Kachru} and in its sequels, the fact of having $%
z\neq1$ introduces new features in the boundary field theory correlators.
For instance, the fact of having replaced AdS space by the Lifshitz space
introduces a dependence of $z $ in the formula that gives the scaling
dimension of the operators in the dual theory: In the case of a scalar field
of mass $m$ in a $D$-dimensional Lifshitz space of dynamical exponent $z$,
the scaling dimension $\Delta$ of the dual operator is given by the formula
\begin{equation}
\Delta^{2}-(D+z-2)\Delta-m^{2}l^{2}=0,  \label{GDelta}
\end{equation}
and its value corresponds to the asymptotic behavior of the bulk field at
large distance, namely%
\begin{equation}
\Phi(x,t,r)\sim r^{-\Delta}\ \Phi_{0}(x,t)+\mathcal{O}(r^{-\Delta -1}),
\label{Gasymptotic}
\end{equation}
with the boundary configuration $\Phi_{0}(x,t)$ acting as a source for a
dual field from the boundary theory point of view. Finiteness of the scalar
field action on (\ref{Gmetric}) demands
\begin{equation}
2\Delta> D+z-2,  \label{Gtorbellino}
\end{equation}
that is, it demands the following $z$-dependent generalization of the
Breitenlohner-Freedman bound
\begin{equation}
m^{2} > -\frac{(D+z-2)^{2}}{4l^{2}}.  \label{GBF}
\end{equation}

\subsection{Correlation functions}

Correlation functions in the dual theory can then be computed by following
the standard holographic prescription, which requires to calculate the
on-shell regularized Euclidean action with appropriate boundary asymptotic.
To do this, one first performs a Wick rotation, defining $\tau=it$. It is
also convenient to define inverse radial coordinate as $u\equiv1/r$, for
which the boundary turns out to be located at $u=0$. Then, one solves the
scalar wave equation on the Euclidean background, whose solution, in the
near boundary region, namely for small values of $u$, behaves like
\begin{equation}
\Phi(x,\tau,u)\sim u^{\Delta_{+}}\ \Phi_{0}^{(+)}(x,\tau)+u^{\Delta_{-}}\
\Phi_{0}^{(-)}(x,\tau),
\end{equation}
where $\Delta_{\pm}$ are the solutions to (\ref{GDelta}), cf. (\ref%
{Gasymptotic}).

To calculate the boundary two-point function one needs the bulk-boundary
propagator. One writes the bulk field configuration in terms of the Green
function $G(x,\tau,u;x^{\prime},\tau^{\prime},0)$; namely
\begin{equation}
\Phi(x,\tau,u)=\int d\tau^{\prime}dx^{\prime}\ \Phi(x^{\prime},\tau^{\prime
},0)\hspace*{0.07cm}G(x,\tau,u;x^{\prime},\tau^{\prime},0).
\end{equation}
In the momentum space the convolution simply reads%
\begin{equation}
\widetilde{\Phi}(k,\omega,u)=\tilde{G}(k,\omega,u)\ \widetilde{\Phi}%
(k,\omega,0).
\end{equation}

Then, one evaluates the bulk action, first integrating by parts and
introducing an infrared cut-off. Namely,
\begin{equation}
S[\Phi]=\int dx\ d\tau\int_{\epsilon}^{\infty}du\hspace*{0.2cm}\left(
-\Phi\partial_{\mu}\left( \sqrt{g}g^{\mu\nu}\partial_{\nu}\Phi\right)
+\partial_{\mu}(\sqrt{g}g^{\mu\nu}\Phi\partial_{\nu}\Phi)\right) =\int dx\
d\tau\left[ \sqrt{g}g^{uu}\Phi\partial_{u}\Phi\right] _{\epsilon }^{\infty}.
\end{equation}
In the momentum space, one finds%
\begin{equation}
S[\Phi]=\int dk\hspace*{0.07cm}d\omega\left( \widetilde{\Phi}(k,\omega ,0)%
\hspace*{0.07cm}\mathcal{F}(k,\omega)\hspace*{0.2cm}\widetilde{\Phi }%
(-k,-\omega,0)\right)
\end{equation}
with the flux function $\mathcal{F}(k,\omega)$ being defined as follows
\begin{equation}
\mathcal{F}(k,\omega)=\left[ \tilde{G}(-k,-\omega,u)\hspace*{0.07cm}\sqrt {g}%
g^{uu}\hspace*{0.07cm}\partial_{u}\tilde{G}(k,\omega,u)\right] _{\epsilon
}^{\infty}.
\end{equation}

This eventually gives the boundary two-point function in the momentum space
\begin{equation}
\mathcal{F}(k,\omega)=\langle O(k,\omega)\hspace*{0.07cm}O(-k,-\omega)\rangle
\end{equation}
where only the flux factor evaluated on $\epsilon\rightarrow0$ contributes,
as the propagator $\tilde{G}(k,\omega,u)$ vanishes when $u\rightarrow\infty$%
. Nevertheless, the quantity one is interested in is the two-point function
in position space, which is obtained by Fourier transforming the expression
above. This requires a careful treatment of the contact terms when Fourier
transforming the distributions \cite{Rusos} and taking the large separation
limit.

In the particular case of strongly coupled $z=2$ Lifshitz points in $D-1=3$
dimensions at zero temperature, it was shown in \cite{Kachru} that the
two-point function in the position space exhibits the form
\begin{equation}
\langle O(x_{1},t_{1})O(x_{2},t_{2})\rangle\simeq\frac{c}{%
|x_{1}-x_{2}|^{2\Delta}}+...\ ,  \label{Gcorrelator}
\end{equation}
where $c$ is a non-vanishing constant, $O(x,t)$ is an operator of the
boundary theory, dual to the bulk operator $\Phi(x,t,r)$, and where the
ellipses stand for contributions that are subleading in the large separation
limit and contact terms. Notice that, in the case $D=4$ with $z=2$ and $m=0$%
, we have $\Delta=4$.

A remarkable feature of (\ref{Gcorrelator}) is that it does not involve
further suppressions coming from overall functions of the scale invariant
quantity $|x_{1}-x_{2}|^{2}/|t_{1}-t_{2}|$ as one could have expected. As
discussed in \cite{Kachru}, such a suppression could have resulted in
ultralocal behavior of the two-point function. Taking a closer look at the
computation of (\ref{Gcorrelator}) one observes that, in general, the reason
why no ultralocal behavior is found in the boundary correlation function, is
the existence of non-analytic contributions at small momenta. In the
position space, such non-analytic contributions yield a non-vanishing
expression when $|x_{1}-x_{2}|>0$. If only positive integer powers of the
momenta were present, then the Fourier transformation would give only
contact terms.

\subsection{Finite temperature and the quasinormal modes}

Finite temperature regime, on the other hand, is also modified with respect
to the standard $z=1$ case. We know from AdS/CFT that adding finite
temperature in the boundary conformal field theory corresponds to
considering a black hole in AdS space. Similarly, in the case of anisotropic
scale invariant theories, adding finite temperature would correspond to
considering in the bulk a black hole solution that asymptotes to (\ref%
{Gmetric}) at large $r$.

The two-point function at finite temperature is computed along the same
lines as in the zero-temperature case: Basically, one solves the scalar
field equation in the Euclidean black hole background imposing
incoming-modes boundary conditions at the horizon. Typically, the
finite-temperature two-point function would exhibit an exponential
damping-off produced by temperature effects. For instance, in $D=4$ and $z=2$
this was explicitly computed in \cite{BalasubramanianMcGreevy}, and the
result was shown to have the form
\begin{equation}
\langle O(x_{1},t_{1})O(x_{2},t_{2})\rangle\simeq\frac{e^{-\sqrt{4\pi T}%
|x_{1}-x_{2}|}}{|x_{1}-x_{2}|^{3/2}}+...\   \label{GcorrelatorA}
\end{equation}

Apart from being necessary to solve the field equation on the black hole
background to compute the two-point function, the black hole quasinormal
mode frequencies also have a direct interpretation from the boundary point
of view:\ Considering a large black hole in the bulk corresponds to
considering a thermal state in the dual boundary theory, and the decay of
the scalar field in the bulk is directly associated to the decay of a
perturbation of such a thermal state, whose relaxation time is given by the
imaginary part of the quasinormal frequency $\omega(k),$ \cite{HH,ThisSachs}.

For the case of $z=2$ Lifshitz black holes, we will explicitly solve the
quasinormal modes, analytically computing the frequencies and then obtaining
the damping-off of the scalar field perturbations. Then, adapting the
computation that holds for asymptotically AdS black holes to the case of
generic $z$, we will also show that the imaginary part of the frequencies
for the scalar field perturbation about the Lifshitz black hole backgrounds
is always negative, proving the stability of these geometries under this
kind of probe perturbations.

\section{Lifshitz black holes}

In this section we will construct a higher-curvature gravity theory that
admits $z=2$ Lifshitz black holes in arbitrary dimension $D$. We begin by
considering the following ansatz
\begin{equation}
ds_{D}^{2}=-\frac{r^{2z}}{l^{2z}}F\left( r\right) dt^{2}+\frac{l^{2}dr^{2}}{%
r^{2}F\left( r\right) }+r^{2}d\vec{x}_{D-2}^{2}\ ,  \label{Metric}
\end{equation}
asking the metric function to obey $F\left( r\rightarrow+\infty\right)
\rightarrow1$ and have a simple root for $r>0$.

Analytic Lifshitz black hole solutions are not easy to be found, and,
consequently, constructing finite temperature gravity duals has so far
required the introduction of strange matter content whose physical
interpretation and theoretical motivation are unclear. An alternative way of
finding such a Lifshitz black hole solution is considering carefully-tuned
higher-curvature modifications to the Einstein-Hilbert gravity action. This
has been done, for instance, in references \cite{AGGH}-\cite{Rn}. Here,
since we will be concerned with the Lifshitz black hole solutions with $z=2$%
, we will present the simplest higher-curvature gravity Lagrangians that
admit asymptotically Lifshitz black holes with $z=2$ in arbitrary dimension $%
D$. As we will see, this amounts to considering terms in the gravity
Lagrangian that are of cubic order in the Riemann tensor.

Then, we look for higher-curvature Lagrangians that admit metric (\ref%
{Metric}) as solutions with $z\neq1$. Since the most conservative
generalization of Einstein gravity in higher dimensions, Lovelock theories,
excludes static Lifshitz black holes \cite{Zegers}, it is necessary to
explore theories of gravity with higher order field equations\footnote{%
The black holes and wormholes in vacuum, found in \cite{DM} and \cite{MaTr}
respectively, exist only for fixed values of the couplings of Lovelock
gravity, at which Birkhoff's theorem are circumvented.}. In such a setup,
black hole solutions have recently been reported; in particular, in the
quadratic $D=3$ massive gravity of \cite{BHT} a black hole with $z=3$ and $%
F\left( r\right) =1-r_{+}^{2}/r^{2}$ was found in reference \cite{AGGH}. In
reference \cite{AGGH2} the existence of such solutions in generic quadratic
gravity theories in $D$ dimensions was explored. However, in the setup of
\cite{AGGH2} it is not possible to achieve the value $z=2$ for the dynamical
exponent, and so a natural next step is to include cubic terms in the
Lagrangian.

We are interested in a theory that would result valid in $D=3$ as well, so
that we consider only higher-curvature terms constructed with the Ricci
tensor $R_{\mu\nu}$. In three dimensions, the Weyl tensor is zero and the
Riemann tensor is fully determined in terms of its trace. Then, we only have
two quadratic invariants, namely ${\mathcal{R}}_{(1)}^{2}=R^{2}$ and ${%
\mathcal{R}}_{(2)}^{2}=R_{\ \nu}^{\mu}R_{\ \mu}^{\nu}$, and three cubic
invariants, ${\mathcal{R}}_{(1)}^{3}=R^{3}$, ${\mathcal{R}}%
_{(2)}^{3}=RR_{\mu\nu}R^{\mu\nu}$, and ${\mathcal{R}}_{(3)}^{3}=R_{\
\nu}^{\mu}R_{\ \rho}^{\nu}R_{\ \mu}^{\rho}$. For $D\geq7$, there are eight
linearly independent, cubic scalars which can be constructed out of three
Riemann tensors without involving its derivatives (see e.g. \cite{Fulling}).
Here, for simplicity, we consider the simplest model defined by the action
\begin{equation}
I=\int d^{D}x\sqrt{-g}\left( \sigma R-2\Lambda+\sum_{i=1}^{2}b_{i}{\mathcal{R%
}}_{(i)}^{2}+\sum_{j=1}^{3}c_{i}{\mathcal{R}}_{(j)}^{3}\right)
\label{action}
\end{equation}

Already in $D=3$ one observes that this Lagrangian admits Lifshitz black
hole solutions with $z=2$. More generally, one verifies that this Lagrangian
happens to admit asymptotically Lifshitz black holes of the form (\ref%
{Metric}) with $z=2$ for all values of $D$. Notice also that, since the
action contains only traces of the Riemann curvature, this theory admits all
Einstein manifolds as solutions, for which $R_{\mu\nu}=\lambda g_{\mu\nu}$
with $\lambda$ being fixed in terms of the couplings. It is not hard to show
that the field equations coming from (\ref{action}) admit the following
black hole metric as a solution
\begin{equation}
ds_{D}^{2}=-\frac{r^{4}}{l^{4}}\left( 1-\frac{r_{+}^{2}}{r^{2}}\right)
dt^{2}+\frac{l^{2}dr^{2}}{\left( r^{2}-r_{+}^{2} \right) }+r^{2}d\vec {x}%
_{D-2}^{2}\ ,  \label{BHmetric}
\end{equation}
provided the coupling constants $\Lambda, \ b_{i},$ and $c_{i}$ obey
specific relations that we write in the Appendix. Here, the inverse of the
Newton constant $\sigma$ and the parameter $l$ can be regarded as two
arbitrary parameters. Theory (\ref{action}) also admits Lifshitz black hole
solutions with other values of $z$.

Black holes (\ref{BHmetric}) present a Hawking temperature $%
T=r_{+}^{2}/(2\pi l^{3})$. In the metric, $r_{+}$ is an integration constant
which represents the location of the event horizon, $r=r_{+}$. This
integration constant is an actual physical parameter provided at least one
of the flat directions is compactified, which results in breaking the
scaling symmetry (\ref{Gsymmetry}). On the contrary, if all the flat
directions are non-compact, then $r_{+}$ can be settled to $1$ by a simple
change of coordinates.

Having given a particular example of a theory that admits $z=2$ Lifshitz
black holes in arbitrary dimensions, we can proceed and study the scalar
response on these backgrounds.

\section{Quasinormal modes}

\subsection{Quasinormal modes for $z=2$ black holes}

As pointed out in \cite{BalasubramanianMcGreevy}, the scalar field response
in backgrounds (\ref{BHmetric}) turn out to be exactly solvable in terms of
hypergeometric functions. This is relevant, for instance, to compute
boundary correlation functions. This is precisely the computation done in
reference \cite{BalasubramanianMcGreevy}, where the finite temperature
two-point function was shown not to exhibit ultralocal behavior either.
Here, we are interested in a different problem, namely that of computing
quasinormal mode frequencies on background (\ref{BHmetric}). To solve this,
we consider a massive scalar field on the black hole metric (\ref{BHmetric})
\begin{equation}
\left( \square-m^{2}\right) \Phi=\dfrac{1}{\sqrt{-g}}\partial_{\mu}(\sqrt{-g}%
g^{\mu\nu}\partial_{\nu}\Phi)-m^{2}\Phi=0\ ,  \label{GFX}
\end{equation}
and assume separability, namely
\begin{equation}
\Phi(\vec{x},t,r)=e^{-i\omega t+i\vec{k}\cdot\vec{x}}R\left( r\right) \ ,
\end{equation}
where $\vec{k}$ is in principle an arbitrary vector. Quasinormal modes would
come from imposing boundary conditions for the scalar field configuration at
the horizon and at the boundary: We ask for incoming modes boundary
conditions at $r=r_{+}$ while we ask the field to vanish at $r=\infty$. To
implement these conditions it is useful to consider the following coordinate
transformation
\begin{equation}
y=\frac{r^{2}-r_{+}^{2}}{r^{2}}\ ,
\end{equation}
for which the range $r=[r_{+},+\infty)$ is mapped to $y=[0,1)$. This
coordinate transformation induces a change of variable in the equation for
the radial part of the scalar field $R\left( r\right) \rightarrow R\left(
y\right) $, which now has to fulfill the following equation%
\begin{equation}
\frac{d^{2}R\left( y\right) }{dy^{2}}+\frac{\left( 2+\left( D-4\right)
y\right) }{2\left( 1-y\right) y}\frac{dR\left( y\right) }{dy}+\frac {%
l^{2}\left( \omega^{2}l^{4}+\left( \omega^{2}l^{4}+k^{2}r_{+}^{2}\right)
y^{2}-\left( 2\omega^{2}l^{4}+k^{2}r_{+}^{2}+m^{2}r_{+}^{4}\right) y\right)
}{4r_{+}^{4}y^{2}\left( 1-y\right) ^{2}}R\left( y\right) =0\ ,
\end{equation}
where $k^{2}=\vec{k}\cdot\vec{k}$. This equation can be transformed into a
hypergeometric equation, whose solutions for $i\omega l^{3}/r_{+}^{2}\notin%
\mathbb{Z}$ read
\begin{equation}
R\left( y\right) =y^{\alpha}\left( 1-y\right) ^{\beta}\left[ A_{1}\
{}_{2}F_{1}\left( a,b,c,y\right) +A_{2}\ y^{1-c}\ {}_{2}F_{1}\left(
b-c+1,a-c+1,2-c,y\right) \right] \ ,
\end{equation}
where $A_{1}$ and $A_{2}$ are integration constants, ${}_{2}F_{1}\left(
y\right) $ stands for the hypergeometric function, and%
\begin{equation*}
\alpha=-\frac{i\omega l^{3}}{2r_{+}^{2}}\ \ ,\ \ \ \beta=\frac{D+\sqrt {%
D^{2}+4m^{2}l^{2}}}{4}
\end{equation*}
where the arguments of the hypergeometric function are
\begin{align}
a & =\frac{1}{4r_{+}^{2}}\left[ \left( 2+\sqrt{D^{2}+4m^{2}l^{2}}\right)
r_{+}^{2}-2i\omega l^{3}+\sqrt{\left( D-2\right) ^{2}r_{+}^{4}-4l^{2}\left(
k^{2}r_{+}^{2}+\omega^{2}l^{4}\right) }\right] \ , \\
b & =\frac{1}{4r_{+}^{2}}\left[ \left( 2+\sqrt{D^{2}+4m^{2}l^{2}}\right)
r_{+}^{2}-2i\omega l^{3}-\sqrt{\left( D-2\right) ^{2}r_{+}^{4}-4l^{2}\left(
k^{2}r_{+}^{2}+\omega^{2}l^{4}\right) }\right] \ , \\
c & =1-\frac{i\omega l^{3}}{r_{+}^{2}}\ .
\end{align}

One can use properties of the hypergeometric functions to express the
solution as a linear combination of functions that are regular in other
points. We prefer to use the basis above in which the problem of recognizing
the ingoing mode at the horizon is simpler.

Imposing ingoing boundary condition at the horizon, which in these
coordinates is located at $y=0$, requires $A_{2}$ to vanish, since around
the horizon the scalar field behaves as%
\begin{align}
\Phi\underset{y\rightarrow0}{\sim}A_{1}e^{-i\omega
t}y^{-ig\omega}+A_{2}e^{-i\omega t}y^{ig\omega} \underset{y\rightarrow0}{\sim%
}A_{1}e^{-i\omega\left( t+g\ln y\right) }+A_{2}e^{-i\omega\left( t-g\ln
y\right) }\ ,
\end{align}
where $g:={l^{3}}/{2r_{+}^{2}}$, and where we have used the fact that $%
{}_{2}F_{1}\left( a,b,c,0\right) =1$. Therefore, setting $A_{2}=0$ gives the
desired behavior, namely ingoing modes at the horizon. One is then left with
the following solution for the radial part%
\begin{equation}
R\left( y\right) =A_{1}y^{\alpha}\left( 1-y\right) ^{\beta}
{}_{2}F_{1}\left( a,b,c,y\right) \ .  \label{Rdex1}
\end{equation}

In order to impose a boundary condition at infinity, $y\rightarrow1$, it is
convenient to use the Kummer's identities for the hypergeometric function
and express ${}_{2}F_{1}\left( y\right) $ in terms of combinations of $%
{}_{2}F_{1}\left( 1-y\right) $. After performing this transformation the
function (\ref{Rdex1}) reads%
\begin{align}
R\left( y\right) & = A_{1}y^{\alpha}\left( 1-y\right) ^{\beta}\left[
\xi_{1}\ {}_{2}F_{1}\left( a,b,a+b+1-c,1-y\right) +\xi_{2}\ \left(
1-y\right) ^{c-a-b} {}_{2}F_{1}\left( c-a,c-b,1+c-a-b,1-y\right) \right] \ ,
\label{Rdex2}
\end{align}
with
\begin{equation}
\xi_{1}=\frac{\Gamma\left( c\right) \Gamma\left( c-a-b\right) }{\Gamma\left(
c-a\right) \Gamma\left( c-b\right) } \ \ \text{ and } \ \ \xi_{2}=\frac{%
\Gamma\left( c\right) \Gamma\left( a+b-c\right) }{\Gamma\left( a\right)
\Gamma\left( b\right) }\ .
\end{equation}

Now, we impose Dirichlet boundary conditions at infinity: Expanding the
expression (\ref{Rdex2}) when $y\rightarrow1$, one has the following leading
order contributions coming from each of the terms inside of the square
bracket of\ (\ref{Rdex2})
\begin{equation*}
\Phi\underset{y\rightarrow1}{\sim}\xi_{1}\left( 1-y\right) ^{\beta}+\xi
_{2}\ \left( 1-y\right) ^{\beta+c-a-b}\underset{r\rightarrow+\infty}{\sim }%
\xi_{1}r^{-\Delta_{+}}+\xi_{2}r^{-\Delta_{-}}\ ,
\end{equation*}
with%
\begin{equation}
\Delta_{\pm}:=\frac{D\pm\sqrt{D^{2}+4m^{2}l^{2}}}{2}\ .
\end{equation}
Notice that this gives the solutions to equation (\ref{GDelta}). For $%
m^{2}\geq0$ the contribution given by $r^{-\Delta_{-}}$ diverges at
infinity, and then in order to impose Dirichlet boundary condition $\Phi%
\underset{y\rightarrow1}{\sim}0$, one is forced to impose%
\begin{equation}
\xi_{2}=\frac{\Gamma\left( c\right) \Gamma\left( a+b-c\right) }{\Gamma\left(
a\right) \Gamma\left( b\right) }=0\ .
\end{equation}
This is fulfilled provided $a=-n$ or $b=-n$ with $n\in\mathbb{Z}_{\geq0}$
with $c-a-b\notin\mathbb{Z}$. These conditions give rise to the following
spectrum
\begin{equation}
\omega=-\frac{i}{l^{3}}\frac{l^{2}k^{2}+r_{+}^{2}\left(
D+m^{2}l^{2}+4n\left( n+1\right) +\left( 2n+1\right) \sqrt{D^{2}+4m^{2}l^{2}}%
\right) }{\sqrt{D^{2}+4m^{2}l^{2}}+2\left( 2n+1\right) }\ ,  \label{Gresult}
\end{equation}
where we have reinserted $r_{+}$. In terms of the dimension $\Delta_{\pm}$
this reads
\begin{equation}
\omega=-\frac{i}{l^{3}}\frac{l^{2}k^{2}+r_{+}^{2}\left( \Delta_{-}(\Delta
_{-}-2(2n+1)-D)+2(2n+D)(n+1)\right) }{2(2n+1)+D-2\Delta_{-}}\ ,
\label{GresultD}
\end{equation}
In the particular case $D=4$, the frequencies (\ref{Gresult}) coincide with
frequencies $\omega_{1}$ of reference \cite{Joel}.

At the special points $c-a-b\in\mathbb{Z}$, that is when $\sqrt{%
D^{2}+4m^{2}l^{2}}/2\in\mathbb{Z}$, the solutions of hypergeometric equation
develop logarithmic dependence. For $m^{2}>0$, the frequencies (\ref{Gresult}%
) are purely imaginary and, besides, the imaginary part is strictly
negative. This implies that there are no oscillatory modes and the field is
purely damped. In this case the leading term in the asymptotic behavior of
the scalar field is given by $\Phi\underset{y\rightarrow1}{\sim}%
r^{-\Delta_{+}}$, that we will call strong asymptotic behavior.

Notice that for $-\left( {D}/{2}\right) ^{2}\leq m^{2}l^{2}<0$ both $%
\Delta_{\pm}>0$, and so both branches $r^{-\Delta_{\pm}}$, fulfill the
Dirichlet boundary condition with a continuos spectrum. This is similar to
the case of AdS spacetime \cite{BSS} for $m_{BF}^{2}l^{2}\leq m^{2}l^{2}<0$,
where $m_{BF}^{2}:=-({D-1})^{2}/({2l})^{2}$ stands for the
Breitenlohner-Freedman mass \cite{BF}. Notice also that the bound $-\left( {D%
}/{2}\right) ^{2}\leq m^{2}l^{2}$ in the asymptotically Lifshitz black hole
with $z=2$, corresponds to the AdS-Breitenlohner-Freedman bound in dimension
$D+1$.

\subsection{Scalar field and Lifshitz soliton}

In three dimensions, the metric (\ref{BHmetric}) allows one to construct a
soliton by performing the following Wick rotation both in the spatial and
the time coordinates; namely%
\begin{equation}
x\rightarrow\frac{il}{r_{+}}\tilde{t}\text{ and }t\rightarrow\frac{il^{3}}{%
r_{+}^{2}}X\ .  \label{dwr}
\end{equation}
The change in the radial coordinated $r=r_{+}\cosh\rho$ in the three
dimensional version of (\ref{BHmetric}), supplemented by the double Wick
rotation (\ref{dwr}), gives the soliton metric%
\begin{equation}
ds^{2}=l^{2}\left[ -\cosh^{2}\rho\ d\tilde{t}^{2}+d\rho^{2}+\cosh^{2}\rho\
\sinh^{2}\rho\ dX^{2}\right] \ ,  \label{soliton}
\end{equation}
where $\rho\geq0$, and the horizon of the black hole is mapped to the center
of the everywhere regular soliton. Asymptotically, this metric has an
anisotropic scaling symmetry with $z=z_{soliton}=1/2$, and its Euclidean
continuation is diffeomorphic to the Euclidean continuation of the black
hole. Given the smoothness of the solution (\ref{soliton}) and its lack of
integration constants, it would be natural to regard this spacetime as a
groundstate. Some of the implications of this in the case $z=3$ have been
already explored in \cite{HTT}, where it was shown that considering as a
background the soliton constructed out from the $z=3$ black hole in
three-dimensional massive gravity, it is possible to give account of the
entropy of the black hole by means of a generalized version of Cardy's
formula.

It is therefore natural to raise the question about the stability of the
solitons constructed in this manner. In particular, if one considers a
scalar field perturbations on the soliton (\ref{soliton}), it is possible to
integrate the field configurations in terms of hypergeometric functions.
Then, by imposing regularity at the origin ($\rho=0$) and vanishing field at
infinity ($\rho\rightarrow\infty$), one finds the normal modes of the field
on the soliton. Nevertheless it is possible to avoid this computation since
requiring ingoing boundary condition at the horizon of the $z=2$ black hole (%
\ref{BHmetric}) in three dimensions corresponds to requiring regularity of
the scalar field at the origin of the soliton (\ref{soliton}), and therefore
the normal modes on the soliton can be obtained directly from the double
Wick rotated quasinormal frequencies (\ref{Gresult}) on the black hole in
three dimensions\footnote{%
The same occurs for BTZ where the soliton obtained after the Wick rotation
is AdS$_{3}$.}. This is, performing the changes%
\begin{equation}
k\rightarrow\frac{r_{+}}{l}i\omega_{sol}\ \ \text{ and }\ \ \omega
\rightarrow\frac{r_{+}}{l^{3}}ik_{sol}\ ,
\end{equation}
for $D=3$ in (\ref{Gresult}), and then solving for the frequencies on the
soliton $\omega_{sol}$, one gets the following spectrum%
\begin{equation}
\omega_{sol}=\pm\left( \left( 2n+|k_{sol}|+1\right) \sqrt{9+4m^{2}l^{2}}%
+3+m^{2}l^{2}+4n\left( n+|k_{sol}|+1\right) +2|k_{sol}|\right) ^{1/2}\ ,
\label{freqsol}
\end{equation}
where $n\in\mathbb{Z}_{\geq0}$, and $k_{sol}$ is the momentum along the $X$
direction in (\ref{soliton}).

Note that the normal frequencies in (\ref{freqsol}) are real provided $%
m^{2}>-{9}/({4l^{2}})$, so the stability of the scalar field propagation on
the soliton is guaranteed on this range.

\subsection{Scalar field stability for generic $z$}

We observe from (\ref{Gresult}) that the imaginary part of the frequency is
actually negative, namely ${Im}(\omega)<0$. As we will show below, this
result could have been shown by a more general argument, which comes from
adapting an argument of \cite{HH}, valid for asymptotically AdS$_{D}$ black
holes, to generic values of $z$. More precisely, having ${Im}(\omega)<0$ for
Dirichlet boundary conditions is a generic feature of the asymptotically
Lifshitz black holes of the form (\ref{Metric}) with\footnote{%
The same argument works without assuming the special form $%
F(r)=y=1-r^{2}_{+} / r^{2}$ but merely requiring $F^{\prime}(r)>0$ outside
the horizon.} $F\left( r\right) =1-{r_{+}^{2}}/{r^{2}}$, and for arbitrary
values of the dynamical exponent $z$. To see this, let us consider the
ansatz
\begin{equation}
ds_{D}^{2}=-\frac{r^{2z}}{l^{2z}}F(r)dt^{2}+\frac{l^{2}dr^{2}}{r^{2}F(r)}%
+r^{2}d\Sigma_{D-2}^{2}\ ,  \label{GGmetric}
\end{equation}
with $d\Sigma_{D-2}^{2}$ being the metric on the ($D-2$)-dimensional base
manifold. By a simple change of variables, metric (\ref{GGmetric}) can be
rewritten as follows
\begin{equation}
ds_{D}^{2}=-r^{2z}F\left( r\right) dv^{2}+2r^{z-1}dvdr+r^{2}d\Sigma
_{D-2}^{2}\ ,
\end{equation}
where we have fixed $l^{2}=1$ for short.

Now, consider again the scalar field equation (\ref{GFX}) and propose the
following form for its solution
\begin{equation}
\Phi=e^{-i\omega v}R\left( r\right) Y\left( \Sigma\right) \ ,
\end{equation}
where $Y\left( \Sigma\right) $ is a harmonic function on $\Sigma$, i.e. $%
\tilde{\nabla}_{\Sigma}^{2}Y\left( \Sigma\right) =-k^{2}Y\left(
\Sigma\right) $ where $\tilde{\nabla}_{\Sigma}^{2}$ is the Laplacian
intrinsically defined on $\Sigma_{D-2}$, in our case $\Sigma_{D-2}\subseteq%
\mathbb{R}^{D-2}$. Using this, one obtains%
\begin{equation*}
Y\left( \Sigma\right) \left[ -i\omega r^{1-z}\frac{dR}{dr}+r^{3-D-z}\frac{d}{%
dr}\left( Fr^{D+z-1}\frac{dR}{dr}-i\omega r^{D-2}R\right) -\left(
k^{2}r^{D+z-5}+m^{2}\right) R\right] \ =0.
\end{equation*}

If one further defines $R\left( r\right) =r^{\frac{2-D}{2}}\psi\left(
r\right) $, this equation becomes
\begin{equation}
\frac{d}{dr}\left( \frac{f(r)}{r^{z-1}}\frac{d\psi}{dr}\right) -2i\omega%
\frac{d\psi}{dr}-V(r) \ \psi=0\ ,  \label{EcuSc}
\end{equation}
where $f\left( r\right) =r^{2z-2}\left( r^{2}-1\right) $, with $r_{+} =1$,
and where the effective potential is
\begin{equation}
V(r)=\frac{1}{4r^{z+1}}\left( \left( D-2\right) \left( D-6+2z\right)
r^{2z-2}\left( r^{2}-1\right) +4\left( \left( D-2\right) +m^{2}\right)
r^{2z}+4k^{2}r^{2z-2}\right) \ ,
\end{equation}

Notice that for $z\geq2$ this potential is strictly positive outside the
black hole. Notice also that in the case $z=1$, this potential is positive
for $r\geq1$.

By manipulating equation (\ref{EcuSc}) and integrating by parts one can show
that the following equation holds
\begin{equation}
\int_{r=1}^{\infty}dr\left( \frac{f(r)}{r^{z-1}}|\frac{d\psi}{dr}%
|^{2}+V(r)|\psi|^{2}\right) =-\frac{\left| \omega\right| ^{2}|\psi_{(r=1)}
|^{2}}{{Im}\left( \omega\right) }\ ,  \label{GHH}
\end{equation}
where Dirichlet boundary conditions for the scalar field were imposed at
infinity. Condition (\ref{Gtorbellino}) guarantees the convergence of the
left hand side of (\ref{GHH}). Then, since the left hand side of (\ref{GHH})
is strictly positive, it demands ${Im}\left( \omega\right) <0$, and then we
conclude the stability of the scalar under perturbations respecting these
boundary conditions, provided the generalized Breitenlohner-Freedman bound (%
\ref{GBF}) is obeyed. This analysis includes, in particular, the $z=3$ black
hole solution of three-dimensional massive gravity, whose quasinormal modes
were recently studied in \cite{Bertha}.

\section{Conclusions}

In this paper we have studied scalar field perturbations in asymptotically
Lifshitz black holes with dynamical exponent $z=2$ in arbitrary dimension $D$%
. These solutions appear, for instance, as exact solutions of
higher-curvature theories of gravity, which consist of augmenting
Einstein-Hilbert action with terms that are quadratic and cubic in the
Riemann tensor.

For these black holes, we have explicitly solved the quasinormal modes for
scalar field perturbations that obey suitable boundary conditions. We have
computed the frequencies of these perturbations explicitly, finding that
they are purely imaginary, with negative imaginary part. This implies the
stability of these geometries under scalar field perturbations that respect
the mentioned boundary conditions. The stability of these Lifshitz black
holes under scalar field perturbations, on the other hand, can be
anticipated by a more general argument: By adapting a computation previously
done for asymptotically AdS$_{D}$ black holes, we gave a concise argument
that proves the stability of a more general class of black holes with
arbitrary value of $z$ provided the generalized Breitenlohner-Freedman bound
(BF) is satisfied.

By performing double Wick rotations of the black hole (\ref{BHmetric}), we
have constructed an everywhere regular solitonic solution in three
dimensions. This solution has an asymptotically anisotropic symmetry. We
have obtained the normal frequencies for the scalar field on this metric by
performing a double Wick rotation of the quasinormal spectrum of the black
hole. This is consistent since asking for an ingoing boundary condition for
the scalar field on the horizon of the black hole corresponds to asking for
regularity of the field at the center of the soliton. The spectrum obtained
in this way shows that the scalar field oscillates without any damping
neither exponential grow, provided the generalized BF bound is fulfilled.
This analysis can be extended to higher dimensions, where in order to
construct a soliton one must single out one of the flat directions on the
metric (\ref{BHmetric}), and then apply a double Wick rotation between the
selected coordinate and time. Performing the corresponding Wick rotation on
the quasinormal spectrum (\ref{Gresult}) itself in arbitrary dimensions
gives the normal modes for the scalar perturbation on the soliton with
frequencies that are real provided the generalized BF bound is fulfilled. It
would be interesting to see if the nonlinear instability of AdS under the
collapse of a backreackting scalar field \cite{Bizon}, also appears in the
case of the asymptotically Lifshitz backgrounds.

Before concluding, we would like to emphasize that in this article we have
only considered the quasinormal mode computation for scalar fields, and even
when from the holographic point of view such computation is interesting by
its own right, concluding stability of asymptotically Lifshitz spacetimes
would also require to consider, in particular, spin-2 excitations. Being a
higher-curvature theory, computing spin-2 perturbations of an action like (%
\ref{action}) about an asymptotically Lifshitz geometry is certainly
cumbersome, and since the equations of motion are of fourth order, the
results would in general exhibit ghosts. Still, particular higher-curvature
models like the one introduced in \cite{BHT} could likely result to be free
of ghosts about Lifshitz black hole solutions as well. This is an
interesting question that requires further study.

\bigskip

The work of A.G., J.O. and S.R. is supported by FONDECYT grants 1110167,
11090281, 11110176, and CONICYT 791100027. The work of G.G. and M.L. is
supported by PICT, PIP and UBACyT grants, from ANPCyT, CONICET and UBA,
Argentina. The authors thank Pedro \'{A}lvarez, Eloy Ay\'{o}n-Beato,
Francisco Correa, and Bertha Cuadros-Melgar for comments. G.G. and M.L.
thank the hospitality of Universidad Austral de Chile during their stay.
After our paper appeared in arXives, we were informed by Oscar I. Fuentealba
Murillo and his collaborators that, in an unpublished paper, he and collaborators performed
computations similar to those of our section 3.2.

\section{Appendix}

The coupling constants of action (\ref{action}) for the theory to admit $z=2$
Lifshitz black holes (\ref{Metric}) in arbitrary dimension $D$ are given by
\begin{align}
\Lambda & =-\frac{\sigma l^{-2}}{3C_{D} }%
D(3D^{6}-26D^{5}+107D^{4}-20D^{3}-80D^{2}+368D+208), \\
b_{1} & =\frac{\sigma l^{2}}{C_{D} }(3D^{4}-15D^{3}+6D^{2}+104D+32), \\
b_{2} & =-\frac{\sigma l^{2}}{C_{D} \left( D+2\right) }%
(D-1)(D^{5}-D^{4}-20D^{3}+108D^{2}+208D+64), \\
c_{1} & =-\frac{\sigma l^{4}}{3C_{D} }(3D+2)(D-8)(D-1)^{2}, \\
c_{2} & =\frac{2\sigma l^{4}}{C_{D} }\left( D^{3}-13D^{2}+8D+4\right) , \\
c_{3} & =\frac{8\sigma l^{4}}{3C_{D} }\left( 4D+1\right) ,
\end{align}
where $C_{D} :=D^{6}-8D^{5}+53D^{4}-130D^{3}+124D^{2}+296D+64$, which has no
integer roots.

\end{document}